\documentstyle[a4,12pt,epsf]{article}

\newcommand{\be}{\begin{equation}}
\newcommand{\ee}{\end{equation}}
\newcommand{\bea}{\begin{eqnarray}}
\newcommand{\eea}{\end{eqnarray}}
\newcommand{\non}{\nonumber}
\begin{document}
%-------------------------------------------------------------------------

\def\abstract#1
{\begin{center}{\large Abstract}\end{center}\par #1}
\def\title#1{\begin{center}{#1}\end{center}}
\def\author#1{\begin{center}{#1}\end{center}}
\def\address#1{\begin{center}{\it #1}\end{center}} 

%-------------------------------------------------------------------------

\hfill
\parbox{6cm}{YITP-99-6 TIT/HEP-414/COSMO-91\par}

\hfill
\parbox{6cm}{ February 1999 \par}
\par
\vspace{5mm}

%-------------------------------------------------------------------------
\title{ {\Large 
{\bf Who's Afraid of Naked Singularities? } } \\ 
\vspace{3mm}
{\large 
{--- Probing timelike singularities with finite energy waves --- }}
}
\vskip 8mm
\author{ {\large  
Akihiro ISHIBASHI$
{}^{\dag}$\footnote{E-mail address: akihiro@yukawa.kyoto-u.ac.jp}
and Akio HOSOYA
${}^{\ddag}$\footnote{E-mail address: ahosoya@th.phys.titech.ac.jp}
} }
\vskip 2mm 
\address{${}^{\dag}$%
{\large {\it Yukawa Institute for Theoretical Physics, \\
Kyoto University, \\ 
Sakyo-ku, Kyoto 606-8502, Japan
}}}
\address{${}^{\ddag}$% 
{\large {\it 
Department of Physics, \\ 
Tokyo Institute of Technology, \\
Oh-Okayama Meguro-ku, Tokyo 152-0033, Japan 
}}}       

%-------------------------------------------------------------------------
\vskip 10mm
\abstract{
\noindent
To probe naked spacetime singularities with waves rather than with 
particles we study the well-posedness of initial value problems  
for test scalar fields with finite energy so that 
the natural function space of initial data is the Sobolev space.
In the case of static and conformally static spacetimes we examine 
the essential self-adjointness of the time translation operator 
in the wave equation defined in the Hilbert space. 
For some spacetimes the classical singularity becomes regular 
if probed with waves while stronger classical singularities 
remain singular. If the spacetime is regular when probed with waves 
we may say that the spacetime is ``globally hyperbolic.'' 
}
\vskip 3mm

\noindent
PACS number(s):
04.20C,04.20D,04.20.Dw,04.50.+h,04.62.+v,11.25.-w

%-------------------------------------------------------------------------
\section{Introduction}
\label{sec:intro}
%-------------------------------------------------------------------------

In general relativity a singular spacetime is defined by 
the geodesic incompleteness~\cite{HE}. However, sometimes 
such a definition gives a very weak singularity which seems almost 
harmless from physical point of view. For example, a spacetime 
from which a single point is taken out is a singular spacetime 
because there is a geodesic curve 
which terminates at the point outside of the spacetime with a finite 
affine time. A stronger \lq physical singularity' appears for example 
at the center of a spherically symmetric black hole, 
where the curvature scalar diverges and therefore the resultant infinite 
tidal force will tear off any physical object. 
The classification of singularities is yet under way 
but has not been completed~\cite{CLARKE1}.

The standard definition of a spacetime singularity is physically based 
on a probe with classical point particles. 
In this paper we shall discuss a wave probe of timelike singularities 
which was initiated by Wald~\cite{WALD} and later
developed by Horowitz and Marolf~\cite{HM}. 
The idea of the probe with waves rather than with classical particles 
is motivated by quantum field theory 
because everything should be described by quantum fields. 
The wave may propagate through the would-be singularity 
with a definite and unique way. 
For example, in the case of hydrogen atom the wave function is 
finite at the origin, which is a classical singularity. 
It is known that if the space is geodesically complete the Laplacian
operator has a self-adjoint extension and the extension is unique so that
the wave propagation is well defined. Converse is not always true.
If the geodesic completeness is replaced by the well-posedness 
of initial value problems for test fields the concept of 
the global hyperbolicity and therefore the cosmic censorship 
%%%%%%%%%%%%%%%%%%%%%%%%%%%%%%%%%%%%
\footnote{
Here we refer to the ``physical formulation'' of the strong 
cosmic censorship rather than the ``precise formulation'' 
in Wald's book~\cite{WaldTEXT}. 
If our wave approach can be extended to the initial value problem of 
the Einstein equations the notion of the cosmic censorship will 
substantially change. 
}
%%%%%%%%%%%%%%%%%%%%%%%%%%%%%%%%%%%%
should be drastically changed as Clarke~\cite{CLARKE2} has advocated.

We shall be concerned with a wave propagation dictated 
by the Klein-Gordon equation in a curved spacetime with timelike 
singularities. Only for illustration in the introduction we use 
the simplest case; the Klein-Gordon equation in $(1+1)$-dimensional 
Minkowski spacetime, 
$ (- \partial^2_t  + \partial^2_x )f =0$ 
defined in a suitable region of the spacetime. 
(For a general case, see the following sections.)

For the initial value problem we introduce the following norm 
on a function space on each $t = const.$ hypersurface: 
\begin{equation}
||f|| := \left( 
\frac{q^2}{2}\int dx |f|^2 
+ \frac{1}{2}\int dx \left| {d{}f\over dx} \right|^2 
\right)^{\frac{1}{2}},
\end{equation}
where $q^2$ is a positive constant. 
We call the function space 
${\cal H} = \{f \; | \; ||f|| < \infty\}$ the {\em Sobolev space} 
or $H^{1}$. The Sobolev norm has been used in the standard formulation 
of well-posed initial value problems in a general globally 
hyperbolic spacetime~\cite{HE}. We note that in general 
the well-posedness of an initial value problem requires 
continuous dependence of solutions on initial data,  
besides the existence and the uniqueness of solutions~\cite{WALD}. 
However, the main issue we will address in this paper 
is to see the uniqueness of solutions of a wave equation in a non-globally 
hyperbolic spacetime, so hereafter we say that 
the initial value problem is well-posed 
when the wave propagation is uniquely determined in the whole 
spacetime.~\footnote{
To show the existence of solutions and to establish an appropriate continuous 
relation between initial data and solutions, Sobolev norms containing higher 
order derivatives are chosen to define a topology on the space of 
initial data. However, to prove the uniqueness of solutions 
of second order linear hyperbolic equations, it is sufficient to adopt $H^1$ 
as our Sobolev space~\cite{HE}, which is larger than $H^{m}$ 
with norms containing $m(>1)$th order derivatives. 
Our results in this paper hold also for $H^{m(>1)}$. 
}

It is known that the norm is bounded above by the field energy so that 
the finiteness of the energy implies the finiteness of the norm~\cite{HE2}.  
Since we cannot afford to prepare an infinite energy field 
configuration as initial data, the function space is naturally 
limited by the condition:
%%%%%%%%%%%%%%%%%%%%%%%%%%%%%%%%%%%%
\footnote{The difference between ours and that of reference~\cite{HM} 
is in the definition of the norm and therefore of the Hilbert space. 
In the case of quantum mechanics the natural Hilbert space is the linear 
function space with the square integrability, $L^2$, because of 
the probabilistic interpretation of the wave function. }
%%%%%%%%%%%%%%%%%%%%%%%%%%%%%%%%%%%%%
\be 
||f|| < \infty. 
\ee 
The corresponding natural inner product in our Hilbert space is defined as
\begin{equation}
(f,g) := \frac{q^2}{2}\int dx f^{*}g+\frac{1}{2}\int dx {d{}f^{*}\over
dx}{dg\over dx}, 
\end{equation}
so that $||f||^2=(f,f)$.

We shall confine ourselves mainly to the case of timelike singularities 
in static or conformally static spacetimes so that the wave equation
becomes of the form,
\be
\partial^2_t \phi = - A \phi,
\ee
where $A$ is an operator which contains spatial coordinates 
and spatial derivatives only.
In this case the well-posedness of the initial value problem 
is translated into the essential self-adjointness of the operator 
$A$ because of the spectral theorem~\cite{RS}. 
Namely, we prepare a smooth and nice initial data at some spatial
hypersurface by choosing the Sobolev space $H^1$ as the Hilbert space. 
The wave will propagate and eventually hit the timelike singularity 
and will be scattered off in someway. The point of the essential 
self-adjointness is that any unwanted singular modes 
which are not contained in the domain of the initial data 
will not appear after scattering so that the initial value problem is
well-posed with no arbitrariness in the choice of the boundary 
conditions and the prediction is unique. 
In such a case we say that the spacetime is ``wave-regular.''

We would like to emphasize the relevance of the present work to 
quantum field theory in curved spacetime. 
It will be natural to expand a quantum field in terms of
the normal modes which belong to the Sobolev space rather than $L^{2}$. 
We assign the coefficients of the mode expansion as annihilation 
and creation operators. 
The quantum states are constructed by applying the creation
operators to the vacuum state which is defined by the condition that the
vacuum is annihilated by all the annihilation operators.~\footnote{ 
The constructed quantum states belong to $L^{2}$ class in the Fock space. 
} 
This construction implies that if the initial value problem 
is well-posed the vacuum expectation value of the energy momentum 
tensor should be well-behaved near the would-be singularities 
so that the field energy is finite.

The organization of the rest of the paper is as follows. 
In Sec.~\ref{sec:hilbert} we propose a natural choice 
of the function space in which the initial value problem 
is explored (Subsec.~\ref{subsec:finite}) 
and we recapitulate the criterion of essential self-adjointness 
of operators in the Hilbert space (Subsec.~\ref{subsec:esa}). 
In Sec.~\ref{sec:removedpt} we demonstrate how we can probe 
singularities with waves in Minkowski spacetime with a single point 
removed and give an intuitive justification of the choice of the Sobolev 
space as the Hilbert space. 
Section~\ref{sec:static} supplies several 
examples of static spacetimes with timelike singularities. 
We explicitly show that many of the classical singularities 
become wave-regular, while a single example is wave-singular. 
Section~\ref{sec:others} is the extensions of the discussion of 
the previous sections to scalar fields with general non-minimal coupling 
and to conformally static spacetimes. 
In Sec.~\ref{sec:Hair} we discuss how to characterize wave-singular 
naked singularities in our approach and propose a notion of hair 
of naked singularities. 
Section~\ref{sec:sumdis} is devoted to summary and discussion. 
In Appendix some mathematical materials on the essentially 
self-adjointness are given for the sake of reader's convenience.

%---------------------------------------------------------------
\section{The function space of initial data}
\label{sec:hilbert}
%---------------------------------------------------------------

%----------------------------------------------
\subsection{Finite energy field configuration}
\label{subsec:finite}
%----------------------------------------------

We consider an $(n+2)$-dimensional static spacetime of the metric form
\be
ds^2 = - V^2 dt^2 + h_{ij}dx^i dx^j, 
\ee
with a timelike Killing vector field $\xi^\mu = (\partial_t)^\mu$.

We choose a function space on each $ t= const.$ hypersurface $\Sigma$ as 
\be 
{\cal H} = \{f \; | \;||f|| < \infty \}
\ee
with the Sobolev norm $||f||$ being given by
\be
||f||^2 := \frac{q^2}{2} \int_\Sigma d \Sigma V^{-1} f^{*}f
+ \frac{1}{2} \int_\Sigma d \Sigma V h^{ij} D_i f^{*} D_j f, 
\label{eq:Sobolev}
\ee
where $q^2$ is a positive constant and $D_i$ denotes the covariant 
derivative with respect to the induced metric $h_{ij}$ on $\Sigma$. 
Here $d\Sigma = d^{n+1} x \sqrt{h}$ is the natural volume element 
on $ \Sigma$. The norm is bounded above 
by a positive constant times the energy integral $E$,
\be
||f||^2 < const. \times E,
\ee
where
\be
E := \int_\Sigma d \Sigma n^\mu \xi^\nu T_{\mu \nu}[f],
\ee
with $n^\mu$ being the unit normal to $\Sigma$. 
Here the energy momentum tensor is given by 
\bea 
T_{\mu \nu} [f] := \frac{1}{2} 
\left(
\nabla_\mu f^{*} \nabla_\nu f + \nabla_\nu f^{*} \nabla_\mu f \right) 
 - \frac{1}{2} g_{\mu \nu}
\left(
\nabla^\sigma f^{*} \nabla_\sigma f + m^2 f^*f
\right) .
\eea 
For $n^{\mu}= V^{-1}(\partial_{t})^{\mu}$ the energy $E$ is expressed by
\bea 
E = \frac{1}{2} \int_\Sigma d \Sigma 
\left( V^{-1}\partial_{t}f^{*}\partial_{t}f + m^2 V f^* f \right) 
 + \frac{1}{2} \int_\Sigma d \Sigma V h^{ij} D_i f^{*} D_j f,
\label{eq:Energy}
\eea 
which motivated us to choose the norm given by Eq.~(\ref{eq:Sobolev}). 
The finiteness of the norm, $||f|| < \infty$, is required 
because we can prepare only a finite energy configuration of the field. 
%%%%%%%%%%%%%%%%%%%%%%%%%%%%%%%%%%%%%%%%%%%
\footnote{Of course, the converse is not necessarily true. That is,
$||f||^2<\infty$ does not mean that the energy is finite in general.
However, in our present analysis of the spacetime with a timelike
(conformal)
Killing vector, the Sobolev space implies the finiteness of the field
energy.} 
%%%%%%%%%%%%%%%%%%%%%%%%%%%%%%%%%%%%%%%%%%%
This leads us to the Sobolev space as the function space
${\cal H}$ on $\Sigma$. 
The energy $E$ is conserved because the energy momentum tensor 
$T^{\mu \nu}$ satisfies the conservation law:
$\nabla_\nu T^{\mu \nu} = 0$
and $\xi^\mu$ satisfies the Killing equation: ${\cal L}_\xi ds^2 = 0 $. 
Then the inner product is naturally defined by
\be
(f,g) := \frac{q^2}{2} \int_\Sigma d \Sigma V^{-1} f^{*}g
+ \frac{1}{2} \int_\Sigma d \Sigma V h^{ij} D_i f^{*} D_j g.
\label{eq:inpro}
\ee

We will consider the massless 
case only because it is known that the initial value problem 
is well-posed for $m\neq 0$ if it is for $m=0$~\cite{HM}.

%%%%%%%%%%%%%%%%%%%%%%%%%%%%%%%%%%%
\subsection{Uniqueness of the time translation operator} 
%\subsection{Essential self-adjointness of a time translation operator} 
\label{subsec:esa}
%%%%%%%%%%%%%%%%%%%%%%%%%%%%%%%%%%% 

Let us briefly recapitulate the mathematics on the essential
self-adjointness of a linear operator $A$ 
on the Hilbert space ${\cal H}$. 
For precise definitions see Appendix, in which we collect relevant 
mathematical materials.

The wave equation of a massless test scalar field, $\Box \phi = 0$, 
reduces to 
\be
\partial^2_t \phi = - A \phi, 
\label{eq:testesa}
\ee
where $ A := - V D^i V D_i $ is a positive symmetric operator on ${\cal H}$ 
if the domain of $A$ is suitably chosen, e.g., 
$C^\infty_0 (\Sigma)$, a set of smooth functions with compact support 
on $\Sigma$, so that it is dense in ${\cal H}$. 
In other words we see by a simple computation,
\bea 
(Af,g) = (f,Ag) + \int_{ \partial \Sigma } dS^{i} 
                    \bigl\{ (A + q^2)f^{*}V\partial_{i}g 
                - V\partial_{i}f^{*}(A + q^2)g \bigr\}
\eea 
so that $A$ is symmetric if $f,g \in C^\infty_0 (\Sigma)$ and 
therefore the surface term above vanishes. 
In most cases this choice of the domain is not very restrictive.

The domain of $A$ can be further extended by relaxing the boundary 
condition so that the extended domain coincides with the domain 
of its adjoint operator. 
The extended operator in this manner is said to be self-adjoint 
and its eigenvalues are real and positive. 
Then, for each self-adjoint extension $A_E$, the time evolution of the field 
is uniquely given by~\cite{WALD} 
\be 
  \phi(t) = \cos ( A_E^{1/2}t ) \phi(0) 
            +  A_E^{-1/2} \sin ( A_E^{1/2}t ) \dot{\phi}(0), 
\label{eq:solution}
\ee 
with $\phi(0),\; \dot{\phi}(0) \in {\cal D}(A_E)$ being any initial data. 
In this sense the self-adjoint extension $A_E$ is a time translation operator.

If there are many possibilities of the self-adjoint extensions, 
we have to choose one of them by imposing a particular boundary 
condition, which is normally imposed by some physical requirement. 
In the case of naked singularities we do not have any criterion to choose 
the boundary condition. Therefore, if the self-adjoint extension is unique, 
there remains no ambiguity in the choice of the boundary conditions. 
A symmetric operator $A$ which has a unique self-adjoint extension 
is called essentially self-adjoint.

The well-posedness of the initial value problem of Eq.~(\ref{eq:testesa}) 
is now turned into the essential self-adjointness of the operator $A$, 
which can be tested by considering solutions of the equations 
\be
 A^* \psi = \pm i \psi , 
\label{eq:criterion}
\ee
and showing that such solutions do not belong to 
our Hilbert space~\cite{RS}.

%--------------------------------------------------------------------
\section{Space with a single point removed} 
\label{sec:removedpt}
%--------------------------------------------------------------------

%------------------------------------------
\subsection{Solution of the wave equation}
%------------------------------------------

\begin{figure}
 \centerline{\epsfxsize = 5cm \epsfbox{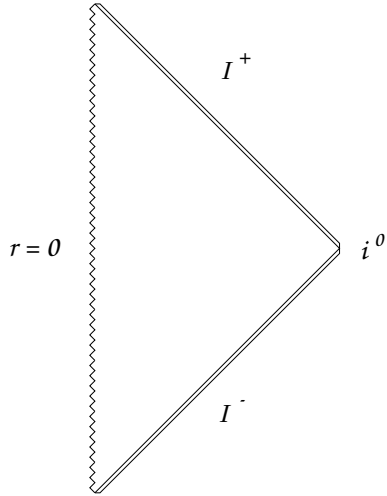}}
\vspace{3mm}
        \caption{A conformal diagram of a spacetime 
         with a timelike singularity at the center.}
        \protect \label{fig:Schwa}
\end{figure}

Let us consider a rather artificial model of a timelike singularity 
which can be fully analyzed. Namely, we consider a spacetime 
which is locally flat but with a single spatial point 
removed so that the spacetime has a timelike singularity 
as illustrated in Fig.~\ref{fig:Schwa} 
and the topology is $({\bf R}^3-\{0\})\times {\bf R}$. 
Our problem in this section is to see the well-posedness of 
the initial value problem of the Klein-Gordon equation, 
\be
   -\partial^2_t \phi + \triangle \phi=0, 
\label{eq:KGr3}
\ee
in this spacetime, which hopefully enhances our understanding 
of the wave probe for more general timelike singularities in the subsequent 
sections and partially supports the choice of the Sobolev space. 
%%%%%%  
%%%%%%
%
%
%\begin{figure}[bht]
%\centerline{\epsfxsize5cm\epsfbox{Schwa.eps}}
%\vspace{3mm}
%\refstepcounter{figure}
%{\small FIG.\thefigure.
%A conformal diagram of a spacetime 
%                 with a timelike singularity at the center.}
%\addcontentsline{lof}{figure}
%{\protect\numberline{\thefigure}{fig1}}
%\label{fig:Schwa}
%\end{figure}
%
%
%%%%%% 

First we assume that our function space ${\cal H}$ on $\Sigma$ is $L^2$, 
i.e., 
\be
{\cal H} = \{\phi \; |\int_{\Sigma} | \phi|^2 d^{3}x < \infty \}, 
\ee
and that the tentative domain ${\cal D}(\triangle)$ of 
the Laplacian $\triangle$ is $C^{\infty}_{0}(\Sigma)$. 
Later we consider the case that $ {\cal H} = H^{1}$ instead of $L^2$ 
and see what is the difference. 
We do not claim that this analysis is new but we demonstrate this 
because we believe that this is the most illustrative explicit model 
in which the choice of the Hilbert space is highlighted. 

Separating the time variable $t$ and the angular 
variables~$\theta, \varphi$ we may write solutions in the form
\be
\phi_{lm} = e^{-ikt} f_{l}(r) Y_{lm}(\theta,\varphi) 
=e^{-ikt}{F_{l}(r)\over r}Y_{lm}(\theta,\varphi)
\ee
with  $Y_{lm}(\theta,\varphi)$ being the spherical harmonics.
The reduced wave equation reads
\be
{d^2 F_{l}\over dr^2}-{l(l+1) F_{l}\over r^2}+k^2 F_{l}=0,
\ee
and the $L^2$ norm squared $\int dx^3 |\phi|^2$ reduces to 
\be
||F||^2=\int_{0}^{\infty}dr |F|^2
\ee
up to an unimportant constant multiple.

The behavior of the radial function near the origin is 
\bea
&\mbox{either}& \;\; {\cal F}_{l}\sim r^{l+1} \quad (f_{l}\sim r^{l}),\\
&\mbox{or}& \;\; {\cal G}_{l} \sim r^{-l} \quad (g_{l}\sim r^{-l-1}). 
\eea 
All the ${\cal F}_{l}$'s belong to the Hilbert space ${\cal H}$. 
The modes ${\cal G}_l \, (l\ge 1)$ are not square integrable at $r=0$ 
and therefore are not normal modes. 
The mode ${\cal G}_{l=0} \sim \mbox{ const.} \, ( g_{l=0} \sim
{\mbox{const.}\over r}$) is the only mode which requires further care.
This mode is square integrable at  $r=0$. In the case of 
$\Sigma \approx {\bf R}^3$ this mode does not belong to our Hilbert 
space because $\triangle ({1\over r})=-4\pi \delta^{3}(x)$ 
is not in $L^2$ class. However, in the case of
$\Sigma \approx {\bf R}^3-\{0\}$ this mode is allowed unless
one further imposes a boundary condition at $r=0$.
However, the boundary condition to be imposed is not unique. 
Actually a boundary condition, 
\be
 a F'- F = 0, 
\label{bc:oblique}
\ee 
is possible at the origin $r=0$, where  $a$ is an arbitrary real parameter. 
In this sort of simple model one can immediately convince oneself that 
this is the most general boundary condition at the origin 
for the self-adjointness of the Laplacian operator 
but there is a systematic way to get the most general boundary condition, 
which is powerful for less simpler cases. 
We defer the demonstration of that method to the following subsections.
Let us concentrate on the S-wave solutions ($l=0$).
The most general S-wave solution which satisfies the above boundary
condition is spanned by
\be
   F_{k}={ \sin(kr)\over k} + a \cdot \cos(kr) 
\ee
with $a$ being the constant in Eq.~(\ref{bc:oblique}).
In this case we say that the self-adjoint extension is not unique 
so that the Laplacian with the initial domain 
${\cal D}(\triangle) = C^{\infty}_{0}({\bf R}^3 - \{0\})$ 
is not essentially self-adjoint and therefore the initial value problem 
is not well-posed. In the case of $\Sigma \approx {\bf R}^3$ we have instead
\be
   F_{k}={ \sin(kr)\over k}, 
\label{eq:Dirichlet}
\ee
which contains no arbitrary parameter so that the Laplacian with 
the domain spanned by $F_k$'s of Eq.~(\ref{eq:Dirichlet}) 
is the only self-adjoint extension.~\footnote{ 
The extended domain spanned by $F_k$'s of Eq.~(\ref{eq:Dirichlet}) 
is the $H^1$ closure of $C^{\infty}_{0}({\bf R}^{3})$, 
and the corresponding self-adjoint extension is 
the Friedrichs extension~\cite{RS}. } 
Therefore the Laplacian with the initial domain $C^{\infty}_{0}({\bf R}^{3})$ 
is essentially self-adjoint.

\begin{figure}
 \centerline{\epsfxsize= 5cm \epsfbox{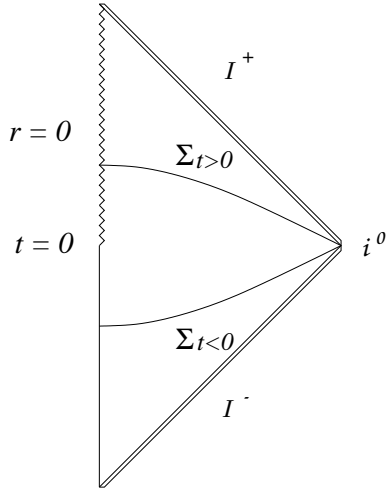}}
\vspace{3mm}
        \caption{ A conformal diagram of a spacetime 
           $M = \Sigma \times {\bf R}$ with 
           $\Sigma_{t<0} \approx {\bf R}^3$ and 
           $\Sigma_{t>0} \approx {\bf R}^3 - \{0 \}$ so 
           that a timelike singularity emerges at the center 
           after $t=0$. }
        \protect \label{fig:Emerge}
\end{figure}

This difference may be slightly more dramatic if we consider 
a spacetime $M = \Sigma \times {\bf R}$: 
\bea 
\Sigma \approx
\left\{
\begin{array}{@{\,}ll}
{\bf R}^{3}-\{0\} 
\quad &\mbox{for}\; t\geq 0,  \\
{\bf R}^{3} \quad & \mbox{for} \; t<0. 
\end{array}
\right. 
\label{eq:Model}
\eea
Namely, a timelike singularity emerges for $t\geq 0$ as depicted 
in Fig.~\ref{fig:Emerge}. 
In such a spacetime the normal modes do not match at $t=0$ unless $a=0$ 
so that the initial value problem is ill-posed. 
%%%%%%  
%%%%%% 
%%%%%%
%
%
%\begin{figure}[bht]
%\centerline{\epsfxsize5cm\epsfbox{Emerge.eps}}
%\vspace{3mm}
%\refstepcounter{figure}
%{\small FIG.\thefigure.
%A conformal diagram of a spacetime 
%           $M = \Sigma \times {\bf R}$ with 
%           $\Sigma_{t<0} \approx {\bf R}^3$ and 
%           $\Sigma_{t>0} \approx {\bf R}^3 - \{0 \}$ so 
%           that a timelike singularity emerges at the center 
%           after $t=0$. }
%\addcontentsline{lof}{figure}
%{\protect\numberline{\thefigure}{fig2}}
%\label{fig:Emerge}
%\end{figure}
%
%
%%%%%% 

%%%%%%%%%%%%%%%%%%%%%%%%%%%%%%%%%%%%%%%%%%%%%%%%%%%%%%%%%
\subsection{A systematic method of self-adjoint extension}
%%%%%%%%%%%%%%%%%%%%%%%%%%%%%%%%%%%%%%%%%%%%%%%%%%%%%%%%%

From the previous subsection we see that the problem of the
function space $L^2({\bf R}^3-\{0\})$ for the field $\phi$ 
reduces to the problem of $L^2(0,\infty)$ for the reduced radial 
wave function $F$. 

Let us study the solutions $F_\pm \in L^2(0,\infty)$ of the equations: 
\be
 - {d^2 F_\pm \over dr^2} = \pm i F_\pm,
\ee
which are reduced from the equations~(\ref{eq:criterion}) 
concentrated on the S-wave again. The solutions are
\bea
{\cal F}_{\pm}&=& \exp \left(-{1\pm i\over \sqrt{2}}r \right)\;, 
\label{eq:F}
\\
{\cal G}_{\pm}&=& \exp \left( {1 \mp i\over \sqrt{2}}r \right). 
\eea
It is clear that the solutions ${\cal G}_{\pm}$ are not 
in $L^2(0,\infty)$ class, while ${\cal F}_\pm$ are.

The prescription to find the most general boundary condition is to compose
\be
F = F_{0}(r)+ {\cal F}_{+}(r) + U {\cal F}_{+}(r),
\ee
where $F_{0}(r)\in C^{\infty}_{0}(0,\infty)$ satisfies the boundary
condition $F_{0}(0)= F'_{0}(0)=0$ at the origin. 
$U$ is the isometry of the space $\{{\cal F}_+ \}$ 
into the space $\{ {\cal F}_- \}$ with respect to the $L^2(0,\infty)$ 
norm, i.e., $U {\cal F}_{+}(r) = e^{i\alpha} {\cal F}_{-}(r)$. 
An elementary computation shows that
\be
{F'(0) \over F(0)}={-{{1+i\over \sqrt{2}} 
- e^{i\alpha} {1-i\over \sqrt{2}}}\over 1 + e^{i\alpha }}
\ee
is a real number which we set equal to $a^{-1}$.
This is what we alluded before.

%%%%%%%%%%%%%%%%%%%%%%%
\subsection{Sobolev space instead of $L^2$}
%%%%%%%%%%%%%%%%%%%%%%% 

Let us now change the Hilbert space from $L^2({\bf R}^3-\{0\})$ space 
to the Sobolev space $H^1({\bf R}^3-\{0\})$. 
We shall look for the solutions for which the integral 
\be
\int_{0}^{\infty}dr r^2 \left| f_\pm'(r) \right|^2 
= \int_{0}^{\infty}dr r^{2} \left|{ \frac{d}{dr} 
\left( {F_\pm(r)\over r} \right) } \right|^2 
\label{eq:int}
\ee
is convergent. However, we can see from Eq.~(\ref{eq:F}) that 
the integral~(\ref{eq:int}) is divergent for ${\cal F}_{\pm}$. 
Therefore $f_{\pm}$ do not belong to the Sobolev space 
$H^1({\bf R}^3-\{0\})$ 
so that there remains no room to relax our boundary condition. 
That is, the Laplacian operator with the initial domain 
$C^\infty_0({\bf R}^3-\{0\})$ is essentially self-adjoint.

Consider now the previous spacetime model: $M = \Sigma \times {\bf R}
$~(\ref{eq:Model}) and the solutions of the field equation~(\ref{eq:KGr3})
in the Sobolev space $H^1(\Sigma)$.
It is now clear that the self-adjointly extended domain of the Laplacian
agrees in both regions of the spacetime.
Therefore the initial value problem is well-posed in the whole spacetime $M$. 
Actually the spherical wave propagates with no trace of 
the would-be singularity at the origin. Of course, this is because the
spacetime is almost Minkowski. 
In a general wave-regular spacetime, the wave would be distorted and 
scattered by strong curvature there in a definite and unique way.

It is physically assuring to see that the removed point 
is completely of no effect if the initial field configuration 
has a finite energy. 
This also supports that our choice of the Hilbert space 
is physically sensible.

%--------------------------------------------------------------
\section{Wave probe in static spacetimes} 
\label{sec:static}
%---------------------------------------------------------------

%%%%%%%%%%%%%%%%%%%%%%%%%%%%%%%%%%%%%%%%
 \subsection{Spherically symmetric static spacetimes}
%%%%%%%%%%%%%%%%%%%%%%%%%%%%%%%%%%%%%%%%

To illustrate the test of the essential self-adjointness of
the operator $A$ in Eq.~(\ref{eq:testesa}) in curved spacetimes, 
we first study the well-known spherically symmetric spacetimes.

In a general $(n+2)$-dimensional spherically symmetric static spacetime, 
the metric is given by 
\be
ds^2 = - V^2 dt^2 + V^{-2} dr^2 + R^2 d\Omega_n. 
\label{eq:n2metric}
\ee
Here we assume that $V^2$ is a positive function of $r$ for $0<r<\infty$
and is singular at $r=0$ so that the causal structure of the spacetime 
is as shown in Fig.~\ref{fig:Schwa}. Provided $\psi = f(r)Y(\Omega)$, 
the equations~(\ref{eq:criterion}) reduce to 
\be
f'' + \frac{(V^2R^n)'}{V^2R^n}f' - \frac{c}{V^2R^2}f 
\pm i \frac{f}{V^4} = 0,
\label{eq:cri_sphe}
\ee
where the prime denotes the derivative with respect to $r$ and 
$c$ is the angular momentum quantum number. 
The norm of $f$ is given by 
\be
||f||^2 = \frac{q^2}{2}\int d\mu dr R^n V^{-2}|f|^2
+ \frac{1}{2}\int d\mu dr R^n V^2|f'|^2, 
\label{eq:norm2}
\ee
where $d\mu$ is the volume element on the unit $n$-sphere 
($d\Sigma = d\mu dr V^{-1}R^n$).

For the essential self-adjointness of $A$, the norms of 
the solutions of the equations~(\ref{eq:cri_sphe}) should be divergent 
for each $c$ and each sign of the imaginary term. 
We can easily verify that the norm $|| f || $ is divergent for $c>0$ 
if it is for $c=0$, so we will examine the essential 
self-adjointness for the $c=0$ (S-wave) case.

%---------------------------------------------------
\subsubsection{Negative mass Schwarzschild spacetime}
%---------------------------------------------------

The $4$-dimensional negative mass Schwarzschild metric is given by 
\be 
V^2 = 1 + \frac{2M}{r}, \quad  R = r, \quad (M>0), 
\ee 
and a timelike singularity is located at the center $r=0$. 
Near the singularity, the equations~(\ref{eq:cri_sphe}) become 
\be 
f'' + \frac{1}{r} f' = 0, 
\label{eq:cri-Schwa}
\ee
since the other terms are less singular or even regular at $r=0$. 
Then the two independent solutions $f= const. $ and $g= \ln r$ are 
obtained. For the latter solution, the second term of 
the norm squared~(\ref{eq:norm2}) behaves as 
\be
\sim \int_{0} dr r^2 V^2 |g'|^2 \sim \ln r|_{0} \rightarrow \infty. 
\ee
Thus the operator $A$ on this spacetime is essentially self-adjoint, 
hence the spacetime is wave-regular.

One might worry that our analysis does not involve the $\pm i$ part 
and only one of the two solutions for Eq.~(\ref{eq:cri-Schwa}) is 
verified not to be in the Hilbert space. 
One may be unhappy about the lack of intuition in the test being 
not completely convinced by the demonstration 
in Sec.~\ref{sec:removedpt}. 
Here, we should remark that the solution $f$ which well behaves 
near the singularity $r=0$ is divergent at infinity $r=\infty$ 
so that the Sobolev norm is divergent. 
The point is that if one of the independent solutions fails to
well behave near the singularity there are no ways for the other solution
to meet the condition at infinity because there is no available other 
independent solution to superpose.

It is amusing to note that this also holds for the higher 
dimensional~($n\geq 3$) negative mass Schwarzschild spacetimes.  
Note also that if $L^{2}$ were chosen as the Hilbert space, 
the operator $A$ would not be essentially 
self-adjoint in this case~\cite{HM}. 

%--------------------------------------------
\subsubsection{Reissner-Nordstr\"{o}m spacetime}
%--------------------------------------------

For the $4$-dimensional over extreme Reissner-Nordstr\"{o}m metric, 
\be  
V^2 = 1 - \frac{2M}{r} + \frac{Q^2}{r^2}, \quad R = r,\;(Q^2\geq M^2), 
\ee 
where $Q$ denotes the electric (magnetic) charge. Near the timelike
singularity at $r = 0$, the equations~(\ref{eq:cri_sphe}) for 
the S-wave $(c=0)$ become $f'' = 0$ ($V^2R^2 \sim Q^2$) 
and $f$ behaves as $f\sim r$ or a constant. 
Then, the norm squared~(\ref{eq:norm2}) is finite. 
Thus, the classical singularity remains wave-singular. 
For $ c > 0$ modes, the norms are divergent. 
This implies that only the S-wave can fall into the singularity.

Our analysis is basically local in time so that our wave approach to 
singularity probe can also work in the case that timelike singularities 
are hidden behind horizons by extending the analysis 
in a straightforward way. 

Higher dimensional generalizations of the Reissner-Nordstr\"{o}m solution 
are given by~\cite{TMP,GM} 
\be  
   V^2 = 1 - \frac{C}{r^{n-1}} + \frac{D^2}{r^{2(n-1)}}, \quad R=r, 
\ee 
with the parameters $C$ and $D$ being proportional to 
the mass and the charge, respectively. 
It is remarked that the singularities in the higher 
dimensional~($n \geq 3$, $D^2 \neq 0$) solutions are also wave-singular.

To summarize the two examples above, 
the timelike singularity in the negative mass Schwarzschild
spacetime is wave-regular while that of the over extreme
Reissner-Nordstr\"{o}m spacetime is not.  The tendency that
 the over extreme Reissner-Nordstr\"{o}m spacetime is more singular than
the negative mass Schwarzschild spacetime  sounds natural because the
curvature is more divergent for  the over extreme Reissner-Nordstr\"{o}m
spacetime.

This reminds us of the well-known examples in quantum mechanics: 
the Coulomb potential and the $r^{-2}$ potential problems 
in $3$-dimensional space. 
The former is essentially self-adjoint and the latter is not if the
$r^{-2}$ potential is attractive and too strong~\cite{RS}.

One might guess from the two examples above that the quasi-local 
mass given (in $4$-dimensional case) by 
\be
M_{local} = - {R\over 2} 
\left\{ g^{\mu \nu}(\partial_{\mu} R)(\partial_{\nu} R) - 1 \right\} 
\ee
would be finite as $r\rightarrow 0$ in the wave-regular cases while it is
infinite in wave-singular cases.  However, this is not the case in other
models as we shall see below. We shall propose an intuitive criterion for
the wave-regularity in Subsec.~\ref{subsec:criterion}.

In general, if the metric functions of the metric~(\ref{eq:n2metric}) 
behave as 
\be  
R \sim r^p,\quad V^2 R^n \sim r^k, 
\label{eq:k}
\ee 
near the singularity $r=0$, the equations~(\ref{eq:cri_sphe}) 
for S-wave ($c=0$) become 
\be 
 f'' + \frac{k}{r}f' = 0, 
\label{eq:fkr}
\ee 
under the condition $np > k-1$, which holds for all our examples. 
Then, the solutions are $f= const.$ and 
$g= \ln r \,(k=1)$, or $r^{1-k}\, (k \neq 1)$. 
Since the norm squared~(\ref{eq:norm2}) for the solution $g = r^{1-k}$ 
(and similarly for $g = \ln r$) is estimated as 
\be
 || g ||^2  \sim \int dr r^{-k + 2(np - k + 1)}+ \int dr r^{-k} 
            \sim \int dr r^{-k}, 
\label{eq:g}
\ee 
the singularity turns out to be wave-regular for the case $k \ge 1$. 
The locally flat example discussed in Sec.~\ref{sec:removedpt} is the case 
$k=2$, the $(n+2)$-dimensional negative mass Schwarzschild metric 
is the case $k=1$ and the $(n+2)$-dimensional 
Reissner-Nordstr\"{o}m metric $k= -n +2$.

%--------------------------------------------
\subsection{Other spacetimes}
%--------------------------------------------

We shall consider some less known but hopefully more physical
solutions of the Einstein equations coupled to matter fields. 
The first two solutions below exhibit null naked singularities 
for some parameter regions. We shall remark on the null naked 
singularities from our point of view.

%---------------------------------
\subsubsection{The Wyman solution}
%---------------------------------

The Wyman solution is a static solution of the $4$-dimensional Einstein
equations coupled to a minimally coupled scalar field~\cite{WYMAN}. 
The metric is given by 
\bea
V^2 &=& \left( 1 - \frac{2 \eta}{r} \right)^{m/\eta}
     = \left( \frac{\rho}{\rho + 2 \eta} \right)^{m/\eta}, 
\non \\
R^2 &=& r^2 \left( 1 - \frac{2 \eta}{r} \right)^{1 - m/\eta}
     = \rho^{1 - m/\eta} \left( \rho + 2 \eta \right)^{1 + m/\eta}, 
\eea
where $ \eta = \sqrt{m^2 + \sigma^2}$ with a scalar charge $\sigma$, 
so $m/\eta <1$ and $\rho := r - 2 \eta$. 
A curvature singularity is located at $\rho = 0$. 
Since $V^2 R^2 \sim \rho$, this is the case of 
the metric functions~(\ref{eq:k}) 
with $k=1$ and thus the spacetime is wave-regular.

%---------------------------------------
\subsubsection{Charged dilaton solution}
%---------------------------------------

The $4$-dimensional charged dilaton solution is given by~\cite{GM} 
\bea
V^2 &=& \left(1 - \frac{r_+}{r} \right)
\left( 1 - \frac{r_-}{r} \right)^{\frac{1-a^2}{1+a^2}}, \\
R^2 &=& r^2 \left( 1 - \frac{r_-}{r} \right)^{\frac{2a^2}{1+a^2}}, 
\eea
with $a^{2}$ being a positive parameter in the model.
Consider the extremal case $r_+ = r_-$.
In the case $a^{2} >1 $, the central singularity at $\rho := r - r_+ = 0$
becomes timelike, while it is null for $a^{2}\leq 1$. 
For any $a^2 >1 $, $R^2V^2 \sim \rho^2$ this is the case 
of the metric functions~(\ref{eq:k}) with $k=2$ 
and the singularity is wave-regular. 
The first term of the norm squared~(\ref{eq:g}) diverges for $a^{2}\leq 3$
so that we would reproduce the result in Ref.~\cite{HM} if we chose
the $L^2$ function space as our Hilbert space.

%--------------------------------------------
\subsubsection{String solution}
%--------------------------------------------

The $5$-dimensional string solution given by 
\bea 
ds^2 &=& V^2 (- dt^2 + dz^2 ) + dr^2 + r^2 d\Omega^2_2, \\  
V^2 &=& \left( 1 + \frac{M}{r} \right)^{-1}, 
\eea
has a curvature singularity at the center $r=0$, which corresponds 
to a straight string. 
The operator $A$ which appears in the wave equation 
$ \Box \phi = - V^{-2}(\partial^2_t + A) \phi = 0$ is expressed as 
\be
A = - \partial^2_z
- V^2 \left\{
\partial^2_r
+ \left(
\frac{\partial_r V^2}{V^2} + \frac{2}{r}
\right) \partial_r
- \frac{c}{r^2}
\right\}, 
\ee
where $c$ is the angular momentum quantum number on the unit
$2$-sphere. By the separation of the variables 
$\psi = f(r) e^{ikz}Y(\Omega)$, 
the equations $(A^* \mp i)\psi = 0$ reduce to 
\be
f'' + \frac{3}{r}f' = 0, \quad \mbox{for S-wave}, 
\ee
near $r = 0$. In this region, the solutions are $f = const.$ 
and $g = r^{-2}$, and the norm squared for $g$ 
\be
||g||^2 \sim \int dr r^2 r^{-4} + \int dr r^2 V^2 r^{-6} 
\ee
diverges. Hence the central singularity is wave-regular.

%------------------------------------------------
\subsection{A simple criterion of wave-regularity 
for spherically symmetric cases} 
\label{subsec:criterion}
%------------------------------------------------

We may give an intuitive but not necessarily mathematically rigorous
explanation of the wave-regularity and a simple criterion 
in what case the classical singularity becomes wave-regular. 
Take the example of static and spherically symmetric spacetimes. 
We can see that if we introduce a new radial coordinate $X$ as 
\be
   X: =\int {dr\over R^nV^2}
\ee
the equations in the test of the essential self-adjointness look like
\be
{d^2 f \over dX^2} - c R^{2n-2}V^2 f \pm iR^{2n} f = 0 , 
\ee  
and the essential part of the Sobolev norm is 
\be 
||f||^2 = \int dX \left|{d{}f \over dX } \right|^2.
\ee
Therefore the problem becomes similar to the quantum mechanics in a
semi-infinite region (except the norm) if the variable $X \in (a,\infty)$ 
for $r\in(0,\infty)$ with $a$ being a finite number. 
As is well known in that case the essential self-adjointness becomes 
non-trivial. On the other hand, if the variable $X \in (-\infty,\infty)$
for $r\in(0,\infty)$, such a ``half-space problem'' would not appear.
For the wave-regular case such as the negative mass Schwarzschild metric 
the range of $X$ extends to $-\infty$ as the singularity $r=0$ is approached 
while it is finite for the over extreme Reissner-Nordstr\"{o}m metric, 
which is the wave-singular case. 
Indeed this holds for all the cases given by the metric 
functions~(\ref{eq:k}). 
This may suggest that the variable $X$ in the wave mechanics plays a role 
similar to the affine parameter in the particle mechanics 
and that for a wave in a wave-regular spacetime 
the singularity is effectively infinitely far away.

As a byproduct of the above observation we can see that if $R^2<\infty$
as $r \rightarrow 0$ and the singularity is null, i.e., 
$\int dr V^{-2} \rightarrow \infty$ as $r \rightarrow 0$,
then the singularity is wave-regular because $\int dr V^{- 2} R^{-n} 
\rightarrow \infty$. This can be checked in the Wyman solution
replacing  $\sigma$ by  $i\sigma$ so that the parameter 
${m}/{\eta} > 1$ and therefore the singularity is null,  
though the scalar field becomes a ghost and the model becomes unphysical. 
In the charged dilaton model, the singularity becomes null 
for $a^2\leq 1$ and is wave-regular. 
The wave-regularity of the null singularity was also asserted 
in Ref.~\cite{HM} but with different reasoning.

%----------------------------------------------
\section{Generalizations}
\label{sec:others}
%----------------------------------------------

%------------------------------------------
\subsection{Probes with other fields }
\label{subsec:otherfield}
%------------------------------------------

So far, the examples of timelike singularities have been probed with 
minimally coupled massless scalar fields. One can think of probing
singularities with other fields such as spinor, vector, tensor fields, 
or metric perturbations. 
Our procedure of probing singularity is immediately generalized
to each case by replacing the inner product~(\ref{eq:inpro}), 
hence the norm, in an appropriate way for the probing field.

For example, when probing a timelike singularity 
with a massless scalar field coupled to the scalar curvature,
we should adopt the inner product respecting the stress-tensor
$T^{c}_{\mu \nu}$ for the field~(see Eq.~(3.190) in Ref.~\cite{BD}).
The field equation in the case is $( \Box + \xi {\cal R}) \phi = 0$,
so the operator
\be
A = - V D^i V D_i - \xi V^2 {\cal R}
\ee
should be examined for the essential self-adjointness, where $\xi$
is a numerical factor and ${\cal R}$ the scalar curvature. 
In the case of the Wyman solution, the equations~$(A^* \mp i)\phi = 0$ 
reduce near $\rho = 0$ to
\be
f'' + \frac{1}{\rho} f' + \frac{ \gamma }{\rho^2} f = 0, \quad
(\gamma = \frac{\xi}{2} \frac{\sigma^2}{\eta^2}),
\ee
and the solutions are given by 
$
f = {\cal A} e^{i\sqrt{\gamma} \ln \rho}
+ {\cal B} e^{- i \sqrt{\gamma} \ln \rho}, \,
({\cal A}, \, {\cal B} \in {\bf C}).
$
After some calculation, it is observed that the Sobolev norms for
the solutions logarithmically diverge near $\rho = 0$.
Thus, the singularity of the Wyman solution is also wave-regular 
when probed with the scalar field coupled to the scalar curvature.
As is well known, in the conformally coupled scalar field case, that is
$\xi = (d - 2)/4(d - 1) $ for any spacetime dimension $d$,
the field equation is invariant under the conformal transformations 
of the metric and the field, $ g_{\mu \nu}(x) \rightarrow 
\bar{g}_{\mu \nu}(x) = C^2(x) g_{\mu \nu}(x)$, 
${\phi} \rightarrow \bar{\phi} = C^{(2 - d)/2} \phi$. 
Since $T^{c}_{\mu \nu}$ transforms as 
$\bar{T}^c_{\mu \nu} = C^{ 2 - d } T^c_{\mu \nu}$, 
the corresponding inner product is conformally invariant.
Thus, the calculation is as simple as that in the static case when
singularities in conformally static spacetimes are probed with conformally
coupled scalar fields. This will be seen in the following subsections.

Generalizations to the probes with spin $1/2$ and $1$ fields are similar. 
Explicit expressions of $T_{\mu \nu}$ for such fields are found, 
for example, in Ref.~\cite{BD}.

%------------------------------------------
\subsection{Conformally static spacetimes}
\label{subsec:cnfsttc}
%------------------------------------------

In the previous section, we have examined the would-be naked singularities 
in static spacetimes, which can be regarded as regular if they are 
not detectable by waves, i.e., if the initial value problem for the field 
is well-posed. 
However, physically more interesting cases which can confront 
with the cosmic censorship are such dynamical spacetimes 
that naked singularities emerge after gravitational collapse. 
For example, to get insights into the problem of the final fate of
gravitational collapse, the Tolman-Bondi solution has been extensively 
studied by many people. It has been revealed that for some 
initial data shell-focusing naked singularities can be formed 
as a final product of spherical dust collapse~\cite{CNJD}. 
As the wave approach to fully dynamical cases is still far reaching 
for us at present, we will present a simple model to mimic 
such a dynamical problem on the basis of our study of 
the timelike singularity in conformally static spacetimes.

%---------------------------------------------------------
\subsubsection{Self-similar case}
%---------------------------------------------------------

Dynamical problems such as the formation of naked singularities 
can be made tractable by assuming self-similarity 
and the nature of naked singularities 
have been investigated~\cite{OPLZJD}.
Self-similar spacetimes also have attracted attention in connection with 
the critical behavior in gravitational collapse 
and have been studied in detail by many authors~\cite{Choptuik}. 
Probing timelike singularities especially in self-similar spacetimes
therefore is an interesting issue concerning the cosmic censorship.
A technical advantage of self-similar metrics is that they can be written
in the conformally static form so that we can straightforwardly apply
our procedure developed for the static spacetime 
in the previous sections to the self-similar spacetimes
when probing with conformally coupled scalar fields.

For a massless scalar field, there is a non-static spherically symmetric 
solution discovered by Roberts~\cite{Roberts:1989}. 
As one of the models of naked singularities in self-similar spacetimes,
we will analyze the timelike singularity in the Roberts solution, 
whose metric can be written in the conformally static form, 
\bea
ds^2 &=& e^{2\eta} d\hat{s}^2 =
e^{2\eta} \left\{ - d\eta^2 + d r^2 + R^2(r) d\Omega^2 \right\}, 
\non \\ 
R^2(r) &:=& \frac{1}{4}
\left\{ 1 + p - (1 - p) e^{- 2 r} \right\} ( e^{2 r} - 1 ), 
\eea 
where $p$ is an integration constant and $(\partial_\eta)^\mu$ is
the homothetic vector.
For the value $0 < p <1$, the curvature singularity located at $ r = 0$
becomes timelike and the global structure is identical to that of
the negative mass Schwarzschild spacetime. 
By using this solution the problem of self-similar scalar field collapse
has been discussed~\cite{Roberts:1989,ONTBrady}.

When probing the singularity with a conformally coupled scalar field, 
we can carry out the previous analysis with respect to the static metric 
$d\hat{s}^2$ instead of $ds^2$. Since near $ r = 0$, $R^2 \sim r$, 
this is the case of the metric functions~(\ref{eq:k}) with $k=1$. 
Thus, the norm squared~(\ref{eq:g}) logarithmically diverges 
so that the singularity is wave-regular for the conformally coupled scalar 
field.

%--------------------------------------------------------------------------
\subsubsection{Conformally flat spacetimes with emerging naked singularity }
%--------------------------------------------------------------------------

More physical situations of gravitational collapse require that 
the spacetime contains a regular initial spacelike hypersurface 
on which the collapsing matter has a compact support.

We can construct a spacetime which has a regular initial hypersurface
$\Sigma_{t_0}$ and a timelike singularity being formed
in the future of $\Sigma_{t_0}$.
For example, let us consider a conformally flat metric 
\be
ds^2 = C^2 \left\{ - dt^2 + dr^2 + r^2 d \Omega^2 \right\}, 
\ee
with a conformal factor which behaves near the center $r=0$ as
\bea
C^2 \sim
\left\{
\begin{array}{@{\,}ll}
r^p & \mbox{($t \ge 0$)}, \\ 
r^p + {1} & \mbox{($ t < 0$)}, 
\end{array}
\right. 
\eea
where $p > - 2$ and $ C^2(r=0,t) \rightarrow 0$ sufficiently smoothly
as $t \rightarrow 0$.
Then, the spacetime has a timelike curvature singularity 
at the center $r=0$ for $t \ge 0$ as depicted in Fig.~\ref{fig:Emerge}. 
However, for sufficiently remote past, 
there are regular hypersurfaces $\Sigma_{t}$. Therefore we can take 
an initial regular hypersurface $\Sigma_{t_0}$ at some $t_0 <0 $ 
and construct the Hilbert space on $\Sigma_{t_0}$. 

Since all the hypersurfaces $\Sigma_t$ are isomorphic to the initial
hypersurface $\Sigma_{t_0}$ up to a conformal factor, the Hilbert spaces 
of a conformally coupled scalar field on $\Sigma_{t}$ are the same, 
even when $\Sigma_{t}$ intersect the central singularity. 
Therefore the spacetime is wave-regular for the conformally coupled 
scalar field. 
\footnote{
An everywhere smooth example can be constructed by choosing 
the conformal factor as
\bea
C^2 =
\left\{
\begin{array}{@{\,}ll}
U_{\ge} \times
e^{- 1/{r^2}} & \mbox{( for $t \ge 0$)},  
\non \\ 
U_{<} \times 
( e^{- {1}/{r^2}} + e^{- 1/{t^2}} ) 
& \mbox{( for $ t < 0$)}, 
\end{array}
\right.
\eea
where $U_{\ge}$ and $U_<$, the analytic functions of $r$ and $t$, 
should be chosen so that $ C^2 \rightarrow 1$ as $ r \rightarrow \infty$. 
By taking $U$'s appropriately, we have regular hypersurfaces at $t<0$ 
whose most of portions are flat. }

Timelike singularities of the type examined above turn out to be 
also wave-regular when probed with the Maxwell field since it is conformally 
invariant and Maxwell's equations are reduced to 
that of a massless Klein-Gordon field under a suitable gauge condition.

%-------------------------------------------
\subsection{Cylindrically symmetric case}
%-------------------------------------------

So far, we have studied spherically symmetric examples, whose central 
singularities are thus considered to be point-like. 
Here, as another example, we will probe a singularity of 
a cylindrically symmetric spacetime given by the metric,
\be
ds^2 = - \rho^{2\sigma_1} dt^2
+ \rho^{2\sigma_2} dz^2 + \rho^{2\sigma_3}d\varphi^2 + d\rho^2, 
\ee 
where the parameters $\sigma_i$ satisfy
$\sum_i \sigma_i = \sum_i (\sigma_i )^2 =1 $. 
This metric is known as the timelike Kasner solution 
and describes a cylindrical vacuum spacetime 
with a timelike curvature singularity along the line $\rho = 0$ 
for the parameter values~$|\sigma_i| <1$.

On each hypersurface $\Sigma_t$, the operator $A$ in Eq.~(\ref{eq:testesa}) 
is written as 
\be A = - \rho^{2 \sigma_1}
\left(
\partial^2_\rho + \frac{1}{\rho} \partial_\rho + \frac{1}{\rho^{2
\sigma_2}} \partial^2_z + \frac{1}{\rho^{2 \sigma_3}} \partial^2_\varphi
\right).
\ee
Since $|\sigma_i| <1$, the equations (\ref{eq:criterion}) reduce to 
\be
\left( \partial^2_\rho + \frac{1}{\rho} \partial_\rho \right) \psi = 0, 
\ee
near the singularity $\rho = 0$. In this region, the solution behaves like 
$f = const.$ or $g= \ln \rho$ and the norm squared becomes 
\be
||\psi||^2 \sim \int d\rho \rho^{1 - 2 \sigma_1} |\psi|^2 
+ \int d\rho \rho | \partial_\rho \psi|^2. 
\ee 
For the solution $g= \ln \rho$, the second term logarithmically 
diverges, thus the singularity is wave-regular.

For $ \sigma_i = (0,0,1)$, the spacetime is flat and, 
if the angle coordinate $\varphi$ has a deficit, 
the line $\rho = 0$ becomes a cone singularity, which thus 
expresses a thin cosmic string in a locally flat spacetime. 
Also in this case the spacetime is wave-regular.

%------------------------------------------------------------------------
\section{Hair of wave-singular naked singularities} 
\label{sec:Hair} 
%------------------------------------------------------------------------

In the previous sections, we have seen that most of the timelike 
``singularities'' are wave-regular and that the singularity 
of the Reissner-Nordstr\"{o}m spacetime is 
the only exception in our examples. 
In the wave-singular case, since probing waves feel the existence 
of the singularity in some sense, we naively expect that the waves 
scattered by the singularity will inform us of some feature of 
the singularity. Here, we shall discuss how to characterize wave-singular 
naked singularities.

In the wave-singular case, the symmetric operator $A$ in the 
wave equation has many different self-adjoint extensions. 
As discussed in Subsec.~\ref{subsec:esa}, 
each self-adjoint extension corresponds to a different boundary condition 
at the singularity and accordingly describes a different time evolution 
of the wave for the same initial data. 
In other words, a wave-singular naked singularity has degrees of freedom 
for the possible choice of the time evolution operator. 
The degrees of freedom can be interpreted as the character or 
the {\em hair} of the wave-singular naked singularity. 
Since the self-adjoint extensions are in one-to-one correspondence with 
the set of partial isometries between the deficiency subspaces 
${\cal K_\pm}$ of $A$ (see Appendix for the definition), 
the set of isometries $U$ describes the degrees of freedom. 

In general, when probing timelike singularities in static spacetimes 
with massless Klein-Gordon waves, we will find a positive real symmetric 
operator of the form $A = - V D^iV D_i$ with domain $C^\infty_0(\Sigma)$. 
Then the deficiency indices $n_+,\,n_-$ of $A$ are equal 
and self-adjoint extensions can be made. 
If the naked singularity is wave-singular, then $n_+ = n_- = N \neq 0$ and 
hence the partial isometry $U$ is represented by 
an $N \times N$ unitary matrix $U(N)$. 
Then, we say that the singularity has a $U(N)$ hair.

Let us consider a wave-singular spacetime which has an asymptotically 
flat region and a static region in a neighborhood of the central timelike 
singularity. More precisely, we will consider such a spacetime that 
the metric form is given by Eq.~(\ref{eq:n2metric}) and the metric functions 
behave as Eq.~(\ref{eq:k}) with $k < 1$ near $r = 0$. 
Consider the wave probe with a massless Klein-Gordon wave, in particular 
the S-wave, in this spacetime. 
The solutions of the equations $(A^* \mp i )\psi_\pm = 0 $ near $r = 0$ are 
given by 
\be 
f_{\pm} \sim a_{\pm}r^{1-k} + b_{\pm}, 
\ee
where $ \psi_\pm = f_{\pm}(r) Y(\Omega)$ and $a_{\pm}, \, b_{\pm} $ 
are constants. Both $f_{\pm}$ well behave near $r =0$. 
On the other hand, in the asymptotic region, 
the equations $(A^*  \mp i )\psi_\pm = 0 $ reduce to 
\bea 
   f_\pm'' + {2 \over r} f_\pm' + \left\{ 
                                 \pm i - {l(l + 1) \over r^2 } 
                          \right\}f_\pm = 0, 
\eea 
and the solutions for S-wave $(l = 0)$ are 
$\exp\left\{(1 \mp i)r/\sqrt{2}\right\}/r$ and 
$\exp\left\{(- 1\mp i)r/\sqrt{2}\right\}/r$. 
Clearly the former solutions diverge in the asymptotic region 
and hence do not belong to the Hilbert space 
while the latter span the deficiency subspaces. 
Therefore the deficiency indices for $l=0$ mode are $(1,1)$ 
and the isometry is $U(1)$. Thus, the wave-singular naked singularity of 
this spacetime has a $U(1)$ hair.

More detailed study will be given 
in the future work, in which we shall investigate what might occur 
in such a wave-singular spacetime in quantum field theory.

%--------------------------------------------
\section{Summary and Discussion} 
\label{sec:sumdis}
%--------------------------------------------

We have studied the well-posedness of the initial value problem of 
the scalar wave equation in the case of static and conformally 
static spacetimes with timelike singularities choosing the Sobolev 
space as the natural Hilbert space. 
The physical idea behind the choice is that we can prepare an
initial data only with finite energy.
We have examined in detail the essential self-adjointness of the operator
$A$ in the wave equation defined in the Hilbert space in various models 
of spacetimes which contain timelike singularities in the conventional
sense. In the spacetimes like the negative mass Schwarzschild spacetime 
the classical singularity becomes regular if probed with waves while more 
stronger classical singularities like the over extreme 
Reissner-Nordstr\"{o}m spacetime remain singular. 

We should comment that the wave-regularity of a spacetime does not 
guarantee that the spacetime is physically realizable. 
For example, a negative mass Schwarzschild spacetime is wave-regular but 
allowance of negative mass solutions would make Minkowski spacetime unstable 
as pointed out by Horowitz and Myers~\cite{HMy}. 
Probably there is a physics which rules out the negative mass Schwarzschild 
solution. In contrast, we may say that the over extreme 
Reissner-Nordstr\"{o}m spacetime is unphysical on the ground 
that it is wave-singular.

\begin{figure}
 \centerline{\epsfxsize = 5.5cm \epsfbox{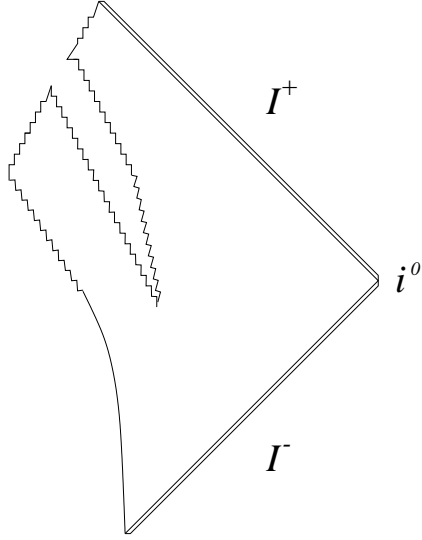}}
\vspace{3mm}
 \caption{A spacetime with emerging would-be naked singularities.}
        \protect \label{fig:Wolf}
\end{figure}

We have briefly touched upon the case that the timelike singularity 
emerges at some point in spacetime in rather artificial models which 
are the Minkowski spacetime with a single spatial point removed 
and the spacetime model which is conformal to Minkowski spacetime.
However, timelike singularities of general spacetimes without any 
(conformal) timelike Killing vector will be more interesting 
from the view point of the cosmic censorship especially in the case 
of gravitational collapse. 
In the case that spacetimes with ``naked singularities'' 
as illustrated by Fig.~\ref{fig:Wolf} are found to be wave-regular, 
such spacetimes can be said to be ``globally hyperbolic" in the sense 
that the initial value problem is well-posed with the unique boundary value 
at the classical singularity. 
Clarke~\cite{CLARKE2} gave a sufficient condition for the well-posedness 
of the initial value problem for test fields. 
However, it turns out that the curve integrability condition 
for a naked singularity in Ref.~\cite{CLARKE2} is not satisfied 
for almost all the wave-regular cases discussed in the present work. 
We suspect that the condition is too demanding for the well-posedness 
of the initial value problem and his theorem can be further sharpened. 

The application of the present work to quantum field theory 
in curved spacetime is most interesting. 
The normal modes are solutions of the wave equation and an analogue of 
the Klein-Gordon inner product exists; 
$i\{ (f,\partial_t g)-(\partial_t f,g) \}$ with $(f,g)$ 
being the inner product providing the Sobolev norm, 
which conserves if the spacetime has a (conformal) timelike Killing vector.

We may carry out similar analyses to spinor, vector, and tensor fields. 
A natural question will be: Does the wave-regularity depend on 
with what fields we probe? 
At the moment what we can say is that in principle yes, 
i.e., it depends and it is nothing wrong from physical point of view. 
If initial value problems are well-posed for {\it all fields}\rm, 
``naked singularities'' are harmless and nothing to be afraid of. 
%%%%%%  
%%%%%% 
%%%%%%
%
%
%\begin{figure}[bht]
%\centerline{\epsfxsize5.5cm\epsfbox{Wolf.eps}}
%\vspace{3mm}
%\refstepcounter{figure}
%{\small FIG.\thefigure.
%A spacetime with emerging would-be naked singularities.}
%\addcontentsline{lof}{figure}
%{\protect\numberline{\thefigure}{fig3}}
%\label{fig:Wolf}
%\end{figure}
%
%
%%%%%% 

\section*{Acknowledgments}

We are very grateful to Dr. H. Kobayashi and Prof. A. Inoue 
for a number of fruitful discussions. 
We also would like to thank Dr. T. Harada for informative comments.  
This work was supported in part by Soryushi Shogakukai (A.I.) 
and also by a grant-in-aid by the Ministry of 
Science, Sports, and Culture of Japan (A.H., 09640341). 

\par 
\vspace*{0.5cm}
%------------------------------------------------------------------------
 \appendix
%------------------------------------------------------------------------

%-------------------------------------------------------------------------
\section*{ Appendix: 
Extensions of symmetric operators} 
\label{app:operator}
%-------------------------------------------------------------------------

We briefly review some definitions of linear operators on a Hilbert space
with an inner product $(\cdot,\cdot)$. 
An operator on a Hilbert space ${\cal H}$ is a pair of a linear mapping 
$ A: {\cal H} \rightarrow {\cal H} $ and its {domain of definition} 
${\cal D}(A)$. 
The pair $(A,{\cal D}(A))$ is often abbreviated by $A$. 
If an operator $A$ with ${\cal D}(A)$ densely defined in ${\cal H}$ 
satisfies 
\be 
(\phi, A\psi) = ( A\phi, \psi), \quad \forall \phi, \psi \in {\cal D}(A), 
\ee 
then $A$ is called {\it symmetric}. 
In the case, any vector $v \in {\cal H}$ can be approximated by vectors 
in ${\cal D}(A)$ as close as possible. 
An operator $A'$ is called an {\em extension} of $A$, 
if ${\cal D}(A) \subset {\cal D}({A}')$ and ${A}'\psi = A \psi, 
\, \forall \psi \in {\cal D}(A)$. 
Extensions of an operator $A$ are obtained by the relaxation of 
the boundary condition on ${\cal D}(A)$. 
Consider sequences $\{\psi_n \} \subset {\cal D}(A)$ such that 
there exist limits 
$\lim_{n \rightarrow \infty} \psi_n =: \xi \in {\cal H}$ 
and $ \lim_{n \rightarrow \infty} A\psi_n =: \zeta \in {\cal H} $. 
If, for every such sequence, $\xi \in {\cal D}(A)$ and $A\xi = \zeta$, 
then $(A,{\cal D}(A))$ is said to be {\em closed}. 
If a non-closed operator $A$ has a closed extension it is called 
{\em closable}. Every closable operator has a smallest closed 
extension, which is called its {\em closure}. 
Consider a symmetric operator $(A, {\cal D}(A))$. 
Define ${\cal D}(A^*)$ to be the set of all $\phi \in {\cal H}$ for which 
there exists $\chi \in {\cal H}$ such that 
\be
(\phi, A\psi) = (\chi, \psi), \quad \forall \psi \in {\cal D}(A). 
\label{eq:ajoint}
\ee
Then, since ${\cal D}(A)$ is dense, $\chi$ is uniquely determined by 
$\phi \in {\cal D}(A^*)$ and Eq.~(\ref{eq:ajoint}). 
An operator $(A^*, {\cal D}(A^*))$ defined by 
$A^* \phi = \chi$ for every $\phi \in {\cal D}(A^*)$ is 
called the {\em adjoint} of $(A, {\cal D}(A))$. 
${\cal D}(A^*)$ may be larger than ${\cal D}(A)$, in which case $A^*$ is 
a proper extension of $A$. 
If $(A^*, {\cal D}(A^*)) = (A, {\cal D}(A))$, 
an operator $(A, {\cal D}(A))$ is said to be {\it self-adjoint}.

Now let us see an example of extensions of symmetric operators to 
self-adjoint ones. Take ${\cal H} = L^2(0,1)$ and consider 
an operator $T$: 
\bea
T\psi &=& - i \frac{d \psi (x) }{dx}, \\ 
{\cal D}(T) &=& \{ \psi \, | \,  \psi(0) = \psi(1) = 0, \,
 \psi \in AC[0,1] \}, 
\eea 
where $AC[0,1]$ expresses the set of absolutely continuous functions 
on $[0,1]$ whose derivatives are in $L^2(0,1)$. 
It can be verified that $T$ is symmetric. 
For $ \phi(x) = \exp(ikx) \in {\cal D}(T^*)$,
$T^* \phi = \chi = k \exp(ikx) \in {\cal H}$ is not an element 
of ${\cal D}(T)$, thus ${\cal D}(T^*) \supset {\cal D}(T)$; 
$T$ is not self-adjoint.

Next, consider an operator $T_{\alpha}$ with the same action 
as $T$ in ${\cal D}(T)$ and with the domain
\be 
{\cal D}(T_{\alpha}) := \{ \psi \, | \, 
\psi(0) = e^{i \alpha} \psi(1), \, \psi \in AC[0,1] \}, 
\ee  
where $\alpha$ is a real number. Clearly, this is an extension of $T$. 
For $\phi \in {\cal D}(T_{\alpha}{}^*)$, 
there exists $\chi \in {\cal H}$ such that 
\be
\forall \psi \in {\cal D}(T_{\alpha}), \quad (\phi, T_{\alpha} \psi) 
= (\chi, \psi), \quad T_{\alpha}{}^* \phi = \chi. 
\ee
Namely,
\be
\int^1_0 dx \phi^* \left( - i \frac{d \psi}{dx} \right) 
= \int^1_0 dx \left( T_{\alpha}{}^* \phi \right)^* \psi.
\ee
On the other hand, by partial integration it is obtained 
\bea
\int^1_0 dx \phi^* \left( - i \frac{d \psi}{dx} \right) 
&=& - i \left[ \phi^* \psi \right]^1_0 
+ \int^1_0 dx \left( T_{\alpha}{}^* \phi \right)^* \psi. 
\eea
Thus, $ \phi^*(1) \psi(1) - \phi^*(0) \psi(0) = 0$. 
From the boundary condition for $\psi$, it follows that 
$ \phi^*(1) = e^{i \alpha} \phi^*(0) = (e^{- i \alpha} \phi(0))^* $, 
hence $ \phi(0) = e^{i \alpha} \phi(1)$. 
Therefore ${\cal D}(T_{\alpha}{}^*) = {\cal D}(T_{\alpha})$; 
$T_\alpha$ is self-adjoint. Since $\alpha$ is arbitrary, it turns out 
that $T$ has infinitely many different self-adjoint extensions.

In general, for a closed symmetric operator $A$, the closed symmetric
extensions can be carried out in more systematic way. 
Consider solutions of $A^* \phi_\pm = \pm i \phi_\pm $ and 
the sets ${\cal K}_\pm$ of the solutions $\phi_\pm$, 
respectively, which are called the {\em deficiency subspaces} of $A$. 
The pair of numbers $(n_+,n_-) := (\dim{\cal K}_+, \dim{\cal K}_-)$ is 
called the {\em deficiency indices} of $A$. 
If $A$ is a closed symmetric operator with the deficiency indices 
$n_+ = n_-$, then $A$ has self-adjoint extensions, 
and if $n_+ = n_- = 0$, $A$ is self-adjoint. 
Let $U$ be the partial isometries ${\cal K}_+ \rightarrow {\cal K}_-$. 
Then, the self-adjoint extensions ${A_E}$ can be obtained 
by taking the domains as 
\be
{\cal D}({A_E}) := \{ \phi_0 + \phi_+ + U \phi_+ \,
| \, \phi_0 \in {\cal D}(A), \, \phi_+ \in {\cal K}_+\}. 
\ee

In the example above, the normalized solutions $\phi_\pm$ 
of the equations $ T^* \phi_\pm = \pm i \phi_\pm $ are 
\be 
\phi_\pm (x) = \frac{\sqrt{2} e^{\mp x}}{\sqrt{\pm (1 - e^{\mp 2})}} . 
\ee 
The deficiency subspaces of $T$ are 
${\cal K}_\pm = \{ \beta \phi_\pm \, | \, \beta \in {\bf C} \}$ and 
thus the deficiency indices are $(1,1)$. 
Then, the partial isometries are taken as 
$U: {\cal K_+} \rightarrow {\cal K_-}: \phi_+ \mapsto \gamma \phi_-$ 
where $| \gamma| = 1$. 
The symmetric extension with respect to $\gamma \in U $ is given by 
$T_\gamma = - i d/dx$ with the domain 
\be 
{\cal D}(T_\gamma) := \{ \phi_o + \beta ( \phi_+ + \gamma \phi_-) \, | \, 
\phi_0 \in {\cal D}(T), \, \beta \in {\bf C} \}. 
\ee 
It turns out that the phase factor $e^{i \alpha}$ is given by 
\be
e^{i \alpha} = \frac{\phi(1)}{\phi(0)} = \frac{1 + \gamma e}{ e + \gamma}. 
\ee

If the closure of a closable symmetric operator $A$ is self-adjoint, 
$A$ is called {\em essentially self-adjoint}. 
In this case, $A$ has a unique self-adjoint extension. 
The basic criterion for essential self-adjointness 
is to verify that its deficiency indices are both zero, namely, 
the solutions $\phi_\pm$ of the equations $(A^* \mp i) \phi_\pm = 0$ 
are not in the considering Hilbert space.
An example of essentially self-adjoint operators is the Laplacian operator
on $L^2 ({\bf R}^n)$ with the domain $C_0^\infty ({\bf R}^n)$: a set of
smooth functions with compact support.

Detailed studies of extensions of symmetric operators and further examples 
can be found in the text book~\cite{RS}.

%%%%%%%%%%%%%%%%%%%%%%%%%%

\end{document}